\documentclass[12pt]{elsart}

\usepackage{graphicx}
\usepackage{subfigure}
\usepackage{amsmath}
\usepackage{amssymb}
\usepackage{cite}

\journal{Physica A}

\begin{document}

\begin{frontmatter}

\title{Mixed spin-1/2 and spin-1 Ising ferromagnets on a triangular lattice}
\author{M. \v{Z}ukovi\v{c}},
\ead{milan.zukovic@upjs.sk}
\author{A. Bob\'{a}k}
\address{Department of Theoretical Physics and Astrophysics, Faculty of Science,\\ 
P. J. \v{S}af\'arik University, Park Angelinum 9, 041 54 Ko\v{s}ice, Slovakia}

\begin{abstract}
By Monte Carlo simulations we study critical properties of the mixed spin-1/2 and spin-1 Ising model on a triangular lattice, considering two different ways of the spin-value distributions on the three sublattices: $(1/2,1/2,1)$ and $(1/2,1,1)$. In the former case, we find standard Ising universality class phase transitions between paramagnetic and magnetic phases but no phase transitions between two distinct magnetic phases $(\pm 1/2,\pm 1/2,\pm 1)$ and $(\pm 1/2,\pm 1/2,0)$ at any finite temperature, except for some interesting non-critical anomalies displayed by response functions. On the other hand, the latter case turns out to be a rare (or perhaps the only) example of a two-dimensional mixed spin-1/2 and spin-1 Ising model on a standard lattice that displays tricritical behavior. 
\end{abstract}


\end{frontmatter}

\section{Introduction}
\hspace*{5mm} Mixed-spin Ising models have been extensively studied by various techniques as simple models of ferrimagnetic and certain types of molecular-based magnetic materials. There are some exact results in special cases~\cite{gonc85,lipo95,jasc98,dakh98,jasc05}, results obtained by mean-field approximation~\cite{kane91,abub01}, effective-field theory with correlations~\cite{kane87,boba97,boba98,kane98a,kane98b,boba00,boba02}, Bethe Peierls method~\cite{iwas84}, cluster variational theory within pair approximation~\cite{tuck01}, renormalization-group calculations~\cite{vero88}, Monte Carlo simulations~\cite{zhan93,buen97,selk00,naka00,naka02,oitm03,godo04,selk10} and other methods~\cite{godo04a,oitm05,oitm06}. These systems have been shown to display variety of interesting critical and compensation properties. The latter refers to the phenomenon observed in some models with ferrimagnetic exchange interaction in which zero total magnetization can be achieved by tuning of temperature below the critical point. The simplest of such lattice models consists of two sublattices one of which is occupied with spins $S=1/2$ and the other with $S=1$. The Hamiltonian of such a mixed-spin model can be written as
\begin{equation}
\label{Hamiltonian}
H=-J\sum_{\langle i,j \rangle}\sigma_{i}S_{j}-D\sum_{j}S_{j}^2,
\end{equation}
where $\sigma_{i}=\pm 1/2$ and $S_{j}=\pm 1,0$ are spins on the $i$th and $j$th lattice sites, respectively, $\langle i,j \rangle$ denotes the sum over nearest neighbors, $J>0$ is a ferromagnetic exchange interaction parameter and $D$ is a single-ion anisotropy parameter. The value of the parameter $D$ encourages either nonmagnetic $S_{j}=0$ ($D<0$) or magnetic $S_{j}=\pm 1$ ($D>0$) states. \\
\hspace*{5mm} Even in the simplest and most studied case of the model on a square lattice, there has been a long standing controversy regarding its critical and compensation behaviors. Only recently an extensive Monte Carlo study has convincingly shown~\cite{selk10} that there are neither tricritical nor compensation points, as had been suggested by some previous approximative approaches~\cite{kane87,kane91,boba97,oitm06}. However, the same study demonstrated the presence of both the tricritical point and a line of compensation points in the three-dimensional model on a simple cubic lattice. This finding might suggest that the increased dimensionality is responsible for the appearance of the tricritical and compensation behaviors. Nevertheless, besides the higher dimensionality, the simple cubic lattice has also higher coordination number $(z=6)$ than the square lattice $(z=4)$. Thus one may ask a question whether the tricritical and compensation phenomena observed in the simple cubic lattice model are exclusive attributes of three-dimensional models or they could also be found in a two-dimensional lattice model with a sufficiently high coordination number. A model on a triangular lattice is an excellent testing example, since it is two-dimensional with the same coordination number ($z=6$) as the three-dimensional simple cubic lattice. We note that both the effective-field~\cite{kane87} and mean-field~\cite{kane91} approximations produced tricritical behavior for a lattice with the coordination number $z=6$. However, these approximative theories do not distinguish between different lattice dimensionalities, i.e. they give the same results for the triangular and simple cubic lattices. Therefore, in this respect, they cannot be trusted, as it was demonstrated on the square lattice~\cite{kane87,kane91}.\\
\begin{figure}[t!]
\centering
\subfigure{\includegraphics[scale=0.45,clip]{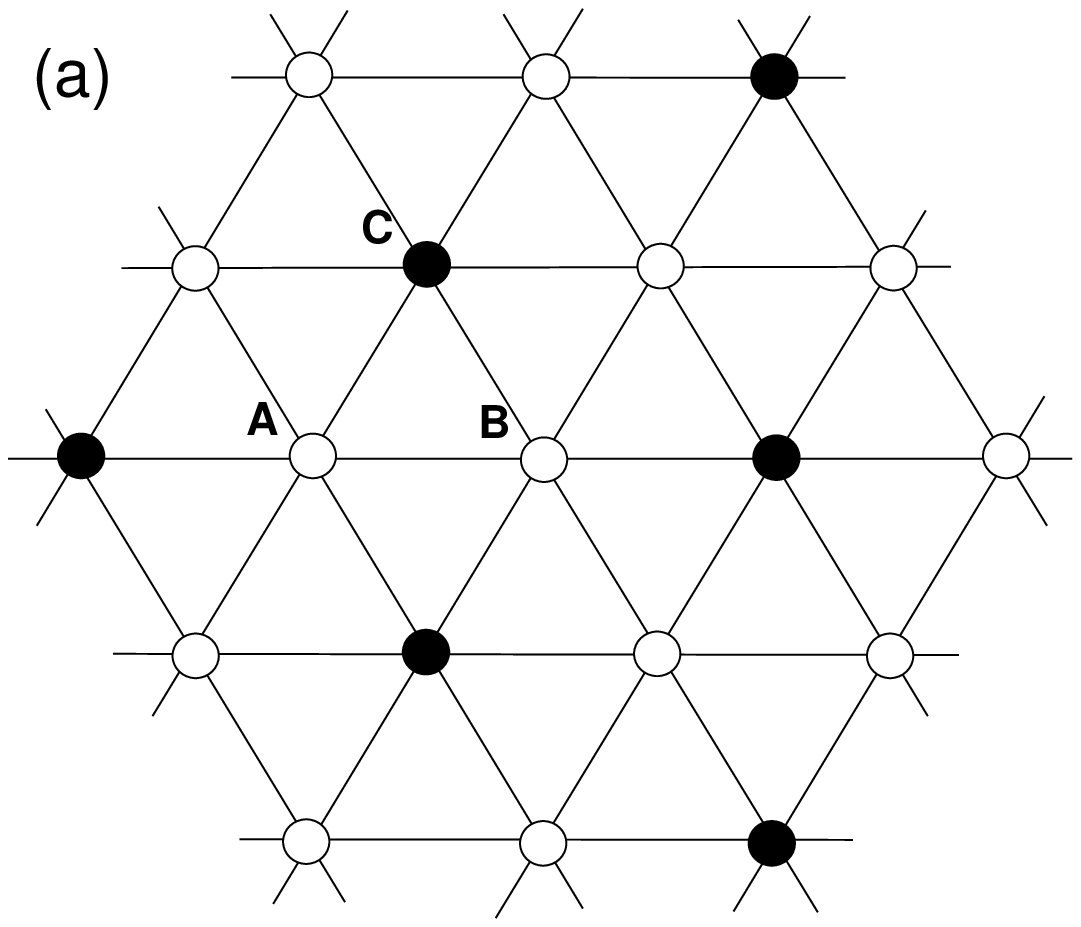}\label{fig:schem_a}}
\subfigure{\includegraphics[scale=0.45,clip]{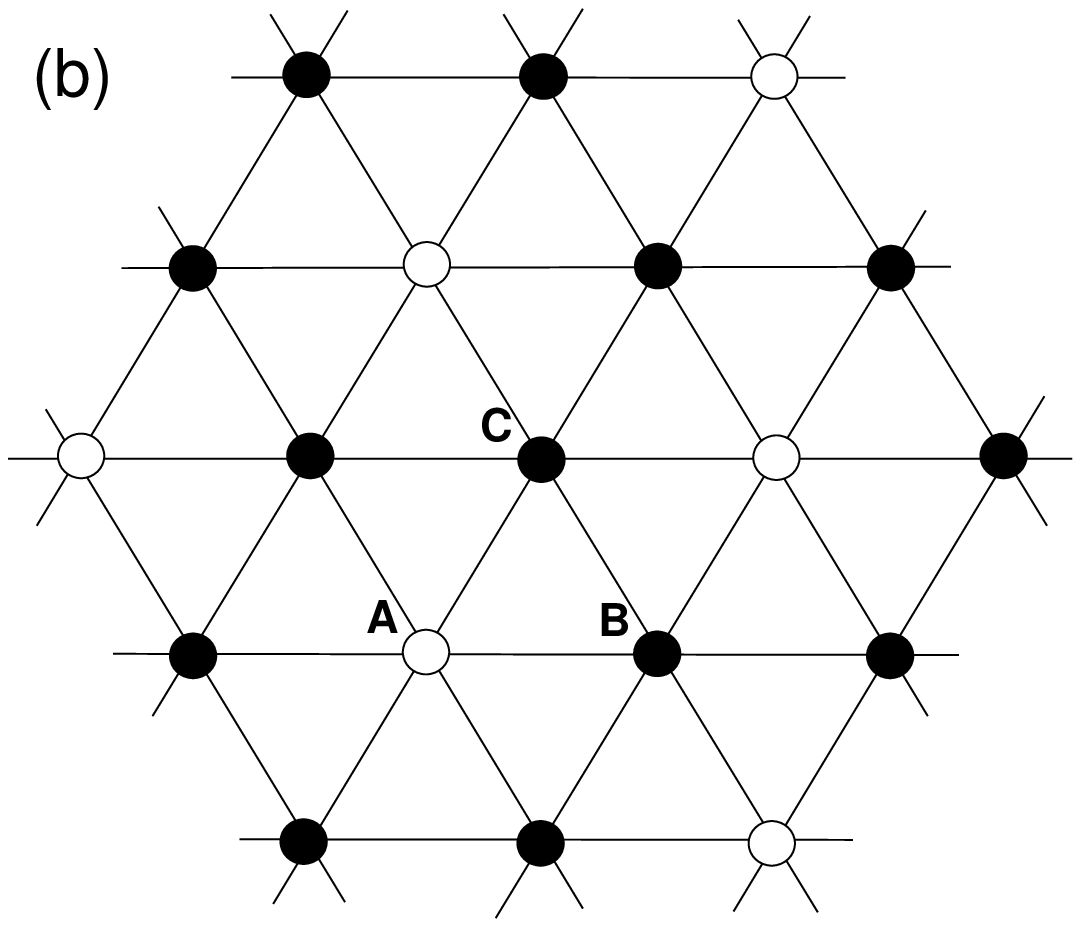}\label{fig:schem_b}}
\caption{Mixed-spin ${\bf S}=(S_{\rm A},S_{\rm B},S_{\rm C})$ models on a triangular lattice consisting of sublattices A, B and C, with (a) ${\bf S}=(1/2,1/2,1)$ mixing and (b) ${\bf S}=(1/2,1,1)$ mixing. Open and solid circles denote spin-1/2 and spin-1 sites, respectively}\label{fig:models}
\end{figure} 
\hspace*{5mm} We point out that the present mixed-spin model is considered on a non-bipartite triangular lattice, consisting of three sublattices A, B and C, occupied with spins ${\bf S}=(S_{\rm A},S_{\rm B},S_{\rm C})$, as shown in Fig.~\ref{fig:models}. This allows to study the model in two mixing modes, unlike in usual bipartite lattices. Therefore, in order to investigate the above described problems, we can consider a mixed-spin ${\bf S}=(1/2,1/2,1)$ model, in which one sublattice is occupied with spin $S=1$ sites and the remaining two sublattices with spin $S=1/2$ sites.
Thus, each spin-1 site is surrounded by $z=6$ nearest neighbors with spin $S=1/2$.  The Hamiltonian for this model can be rewritten in a more specific way as
\begin{equation}
\label{Hamiltonian_I}
H=-J\Big(\sum_{i \in A,j \in B}\sigma_{i}\sigma_{j}+\sum_{i \in A,k \in C}\sigma_{i}S_{k}+\sum_{j \in B,k \in C}\sigma_{j}S_{k}\Big)-D\sum_{k \in C}S_{k}^2.
\end{equation}
An alternative way of spin-mixing is realized in a ${\bf S}=(1/2,1,1)$ model, which is obtained when the spin-1/2 and spin-1 sites in the ${\bf S}=(1/2,1/2,1)$ model are swapped, and in which case the Hamiltonian takes the form
\begin{equation}
\label{Hamiltonian_II}
H=-J\Big(\sum_{i \in A,j \in B}\sigma_{i}S_{j}+\sum_{i \in A,k \in C}\sigma_{i}S_{k}+\sum_{j \in B,k \in C}S_{j}S_{k}\Big)-D\Big(\sum_{j \in B}S_{j}^2+\sum_{k \in C}S_{k}^2\Big).
\end{equation}
Noticing that in each model the two sublattices with the same spins form either spin-1/2 or spin-1 Ising models on a honeycomb backbone, the two models can be anticipated to show qualitatively different critical behaviors. For example, in the large negative $D$ limit the ferromagnetic state is expected to pertain on the spin-1/2 honeycomb backbone of the ${\bf S}=(1/2,1/2,1)$ model, while no ordering can be expected in the ${\bf S}=(1/2,1,1)$ model, when the spin-1 sublattices forming the honeycomb backbone switch to the nonmagnetic states. \\
\hspace*{5mm} Motivated by the quest to find a two-dimensional mixed spin-1/2 and spin-1 Ising model that would show tricritical behavior, which might be realized in the ${\bf S}=(1/2,1,1)$ mixing, and to investigate the nature of the transition between the two ferromagnetic phases $(\pm 1/2,\pm 1/2,\pm 1)$ and $(\pm 1/2,\pm 1/2,0)$, expected to appear in the ${\bf S}=(1/2,1/2,1)$ mixing, we perform Monte Carlo simulations of the above models and determine the respective phase diagrams. 

\section{Monte Carlo simulation}
We employ Monte Carlo (MC) method with the Metropolis dynamics and the periodic boundary conditions. We consider the linear lattice sizes ranging from $L=24$ up to $L=120$. In standard MC simulations, for thermal averaging we typically use $N=2\times 10^5$ up to $10^6$ MCS (Monte Carlo sweeps), after discarding another $20\%$ of MCS needed to bring the system to equilibrium. In order to obtain temperature dependencies at a fixed value of $D$, simulations start from the paramagnetic phase using random initial configurations. Consequently, the temperature is gradually lowered and a new simulation starts from the final configuration obtained at the previous temperature. To obtain variations of the quantities as functions of the single-ion anisotropy parameter $D$, we run simulations at a fixed temperature, which start from appropriately chosen states (i.e., not necessarily random), expected in the considered region of the parameter space. Such an approach ensures that the system is maintained close to the equilibrium in the entire range of the changing parameter and considerably shortens thermalization periods. In order to reliably estimate statistical errors, we applied the $\Gamma$-method~\cite{wolf04}, which has been shown to produce more certain error estimates than the binning techniques. To obtain critical exponents, close to the critical region we perform more extensive simulations using up to $N=10^7$ MCS and apply the reweighing techniques~\cite{ferr88}. The critical exponents are then extracted from the finite-size scaling (FSS) analysis, using the linear sizes $L=24,48,72,96$ and $120$.\\
\hspace*{5mm} We calculate the internal energy per spin $e=\langle H \rangle/L^2$, the respective sublattice magnetizations per site $m_{\rm X}$, (X = A, B or C), as order parameters on the respective sublattices, which for the ${\bf S}=(1/2,1/2,1)$ model are given by
\begin{equation}
\label{subAB_mag}
m_{\rm A(B)} = 3\langle |M_{\rm A(B)}| \rangle/L^2 = 3\Big\langle\Big|\sum_{i \in {\rm A(B)}}\sigma_{i}\Big|\Big\rangle/L^2,
\end{equation}
\begin{equation}
\label{subC_mag}
m_{\rm C} = 3\langle |M_{\rm C}| \rangle/L^2 = 3\Big\langle\Big|\sum_{i \in {\rm C}}S_{i}\Big|\Big\rangle/L^2,
\end{equation}
and for the ${\bf S}=(1/2,1,1)$ model by
\begin{equation}
\label{subA_mag}
m_{\rm A} = 3\langle |M_{\rm A}| \rangle/L^2 = 3\Big\langle\Big|\sum_{i \in {\rm A}}\sigma_{i}\Big|\Big\rangle/L^2,
\end{equation}
\begin{equation}
\label{subBC_mag}
m_{\rm B(C)} = 3\langle |M_{\rm B(C)}| \rangle/L^2 = 3\Big\langle\Big|\sum_{i \in {\rm B(C)}}S_{i}\Big|\Big\rangle/L^2. 
\end{equation}
where $\langle\cdots\rangle$ denotes thermal average. Then, we define the total magnetization per site $m$, as an order parameter of the entire system, for the model ${\bf S}=(1/2,1/2,1)$ given by
\begin{equation}
\label{mag2}
m = \langle |M| \rangle/L^2 = \Big\langle\Big|\sum_{i \in {\rm A}}\sigma_{i}+\sum_{j \in {\rm B}}\sigma_{j}+\sum_{k \in {\rm C}}S_{k}\Big|\Big\rangle/L^2,
\end{equation}
and for the model ${\bf S}=(1/2,1,1)$ by
\begin{equation}
\label{mag1}
m = \langle |M| \rangle/L^2 = \Big\langle\Big|\sum_{i \in {\rm A}}\sigma_{i}+\sum_{j \in {\rm B}}S_{j}+\sum_{k \in {\rm C}}S_{k}\Big|\Big\rangle/L^2.
\end{equation}
Further, we calculate the susceptibilities pertaining to quantities $O=M_{\rm X}$ (X = A, B and C) and $M$ 
\begin{equation}
\label{chi}\chi_{O} = \frac{\langle O^{2} \rangle - \langle O \rangle^{2}}{N_Ok_{B}T}, 
\end{equation}
the specific heat per site $c$
\begin{equation}
\label{c}c=\frac{\langle H^{2} \rangle - \langle H \rangle^{2}}{N_Ok_{B}T^{2}},
\end{equation}
where $N_O$ is the number of sites on the (sub)lattice on which $O$ is defined, and, finally, the logarithmic derivatives of $\langle O \rangle$ and $\langle O^{2} \rangle$ with respect to $\beta=1/k_{B}T$,
\begin{equation}
\label{D1}D1_{O} = \frac{\partial}{\partial \beta}\ln\langle O \rangle = \frac{\langle O H
\rangle}{\langle O \rangle}- \langle H \rangle,
\end{equation}
\begin{equation}
\label{D2}D2_{O} = \frac{\partial}{\partial \beta}\ln\langle O^{2} \rangle = \frac{\langle O^{2} H
\rangle}{\langle O^{2} \rangle}- \langle H \rangle,
\end{equation}
the fourth-order Binder cumulant~\cite{bind81} corresponding to the order parameter $O$
\begin{equation}
\label{eq.U}U_O = 1-\frac{\langle O^{4}\rangle}{3\langle O^{2}\rangle^{2}},
\end{equation}
and finally the fourth-order Binder cumulant for the internal energy
\begin{equation}
\label{eq.V}V = 1-\frac{\langle H^{4}\rangle}{3\langle H^{2}\rangle^{2}}.
\end{equation}
For the FSS analysis we use the following scaling relations, applied to the maximum values of the following functions:
\begin{equation}
\label{scalchi}\chi_{O,max}(L) \propto L^{\gamma_O/\nu_O},
\end{equation}
\begin{equation}
\label{scalD1}D1_{O,max}(L) \propto L^{1/\nu_O},
\end{equation}
\begin{equation}
\label{scalD2}D2_{O,max}(L) \propto L^{1/\nu_O},
\end{equation}
\noindent where $\nu_O$ and $\gamma_O$ are the critical exponents of the correlation length and susceptibility, respectively, pertaining to the quantity $O$. The order parameter cumulant $U_O$ can serve for a simple yet relatively precise location of the phase transition point as a point at which the cumulant curves obtained for different system sizes intersect and at which it achieves a universal value, e.g., $U_O(T_c)=0.611$ for a two-dimensional Ising model~\cite{kame93}. On the other hand, the energy cumulant $V$ is useful to identify a first-order phase transition. In particular, near the transition point it exhibits a minimum the value and position of which scale with the system size $L^{-d}$, where $d$ is the lattice dimension, and in the limit of $L\rightarrow \infty$ it achieves some non-trivial value $V^{*}<2/3$~\cite{chal86,ferr88}.

\section{Results}
\subsection{Ground state}
The triangular lattice system consists of three interpenetrating sublattices A, B and C, as schematically depicted in Fig.~\ref{fig:models}. Focusing on a triangular elementary unit cell consisting of the spins $S_{\rm A}$, $S_{\rm B}$, $S_{\rm C}$, and considering all possible spin states, one can obtain the following phases and expressions for the corresponding reduced ground-state (GS) energies per spin. 
\begin{itemize}
\item Model ${\bf S}=(1/2,1/2,1)$:
	\begin{enumerate}
	\item Phase $(\pm 1/2,\pm 1/2,\pm 1)$ - ferromagnetic state with $S_{\rm A}=S_{\rm B}=\pm\frac{1}{2},S_{\rm C}=\pm 1$ and the energy $e/J=-5/4-\frac{D}{3J}$;
	\item Phase $(\pm 1/2,\pm 1/2,0)$ - ferromagnetic state with $S_{\rm A}=S_{\rm B}=\pm\frac{1}{2},S_{\rm C}=0$ and the energy $e/J=-\frac{1}{4}$.
	\end{enumerate}
\item Model ${\bf S}=(1/2,1,1)$:
	\begin{enumerate}
	\item Phase $(\pm 1/2,\pm 1,\pm 1)$ - ferromagnetic state with $S_{\rm A}=\pm\frac{1}{2},S_{\rm B}=S_{\rm C}=\pm 1$ and the energy $e/J=-2-\frac{2D}{3J}$;
	\item Phase $(0,0,0)$ - nonmagnetic state with $S_{\rm A}=\pm\frac{1}{2}$ (equally in states $+\frac{1}{2}$ and $-\frac{1}{2}$), $S_{\rm B}=S_{\rm C}=0$ and the energy $e/J=0$.
	\end{enumerate}
\end{itemize}
For example, the GS energy of the phase (1) of the model ${\bf S}=(1/2,1/2,1)$ can be obtained from the Hamiltonian~(\ref{Hamiltonian_I}) as follows. Let us consider that all the spin values are positive (all negative values would give the same energy). Spin $\sigma_{\rm A}=1/2$ has three $\sigma_{\rm B}=1/2$ and three $S_{\rm C}=1$ NN, spin $\sigma_{\rm B}=1/2$ has three $\sigma_{\rm A}=1/2$ and three $S_{\rm C}=1$ NN and spin $S_{\rm C}=1$ has three $\sigma_{\rm A}=1/2$ and three $\sigma_{\rm B}=1/2$ NN, so the reduced interaction energy is $-1/6*[1/2*(3*1/2+3*1)+1/2*(3*1/2+3*1)+1*(3*1/2+3*1/2)]=-5/4$. The prefactor 1/6 includes division by two to prevent double counting of NN pairs and by three to obtain the mean value per one spin. The single-ion contribution comes only from the spin $S_{\rm C}=1$ thus it is equal to $-(D/3J)*(0+0+1)$. Then, the total reduced energy per spin is $e/J =-5/4-D/3J$. Ground states for different values of the reduced parameter $D/J$ can be determined by comparing the above energies. By doing so we find that in either model the critical value of $D_c/J=-3$ separates the phase (1) occurring for $D/J>D_c/J$ from the phase (2) occurring for $D/J<D_c/J$.

\subsection{Monte Carlo}
\subsubsection{Model ${\bf S}=(1/2,1/2,1)$}
\begin{figure}[t!]
\centering
\subfigure{\includegraphics[scale=0.48,clip]{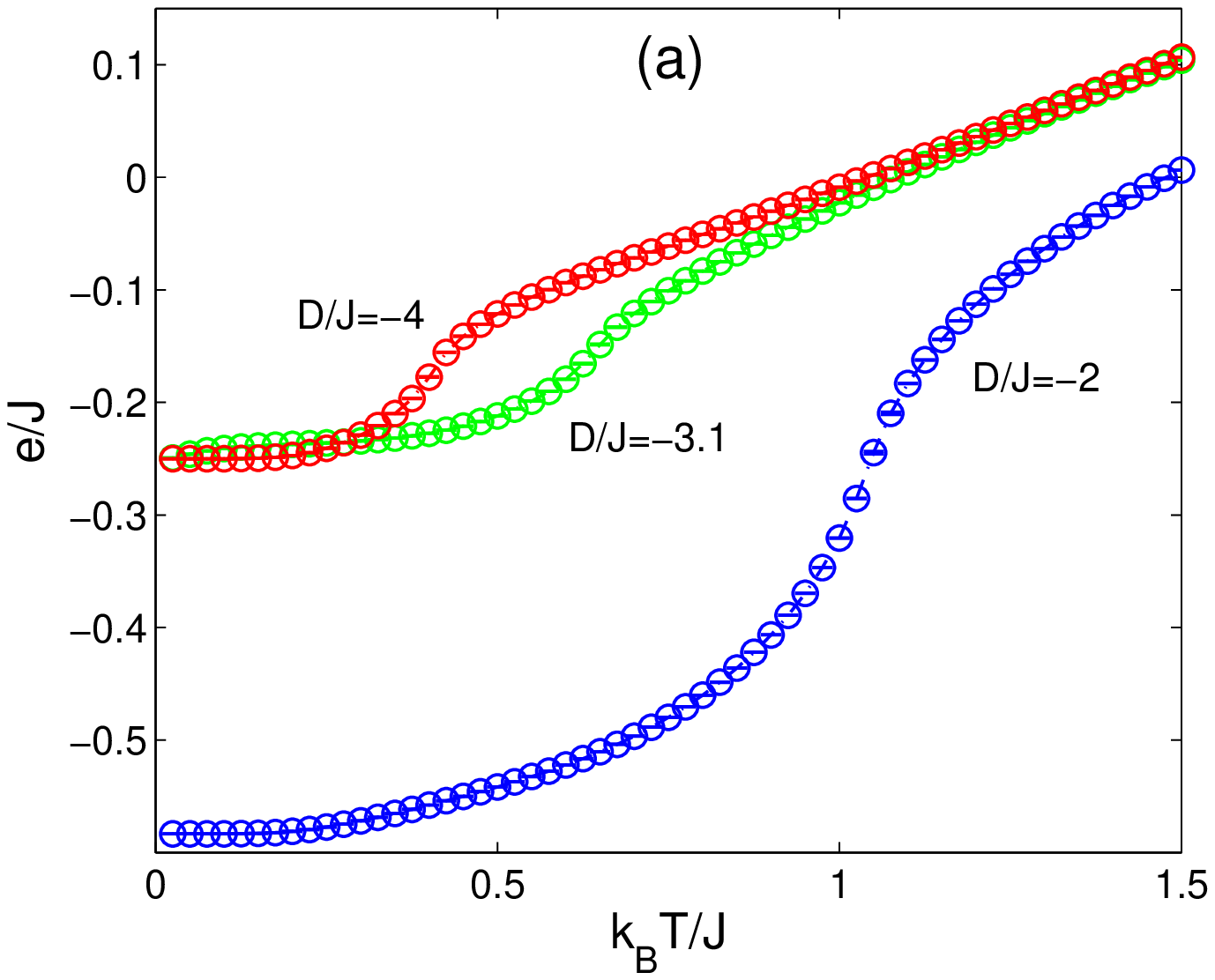}\label{fig:e-T_L48}}
\subfigure{\includegraphics[scale=0.48,clip]{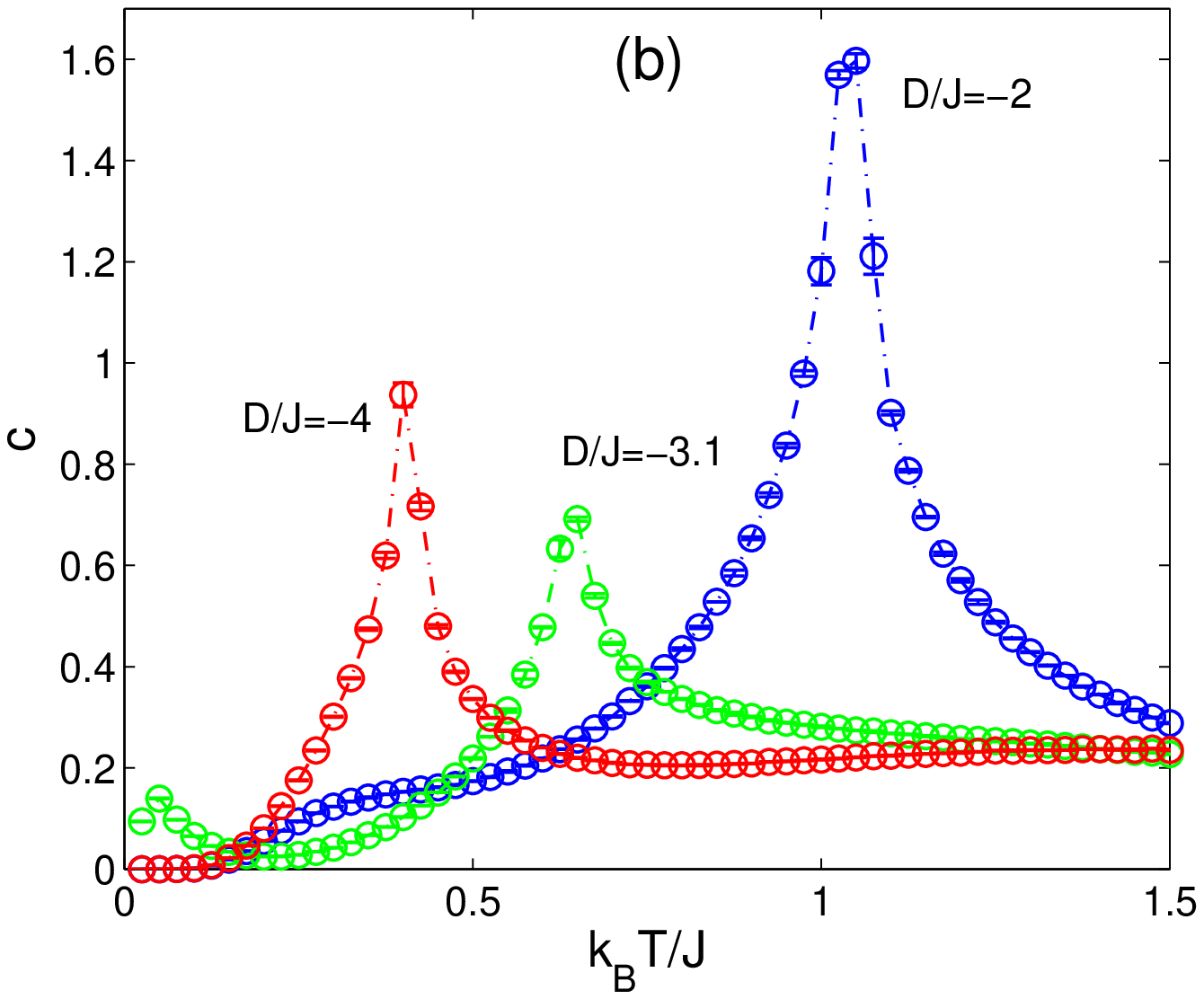}\label{fig:c-T_L48}}
\subfigure{\includegraphics[scale=0.48,clip]{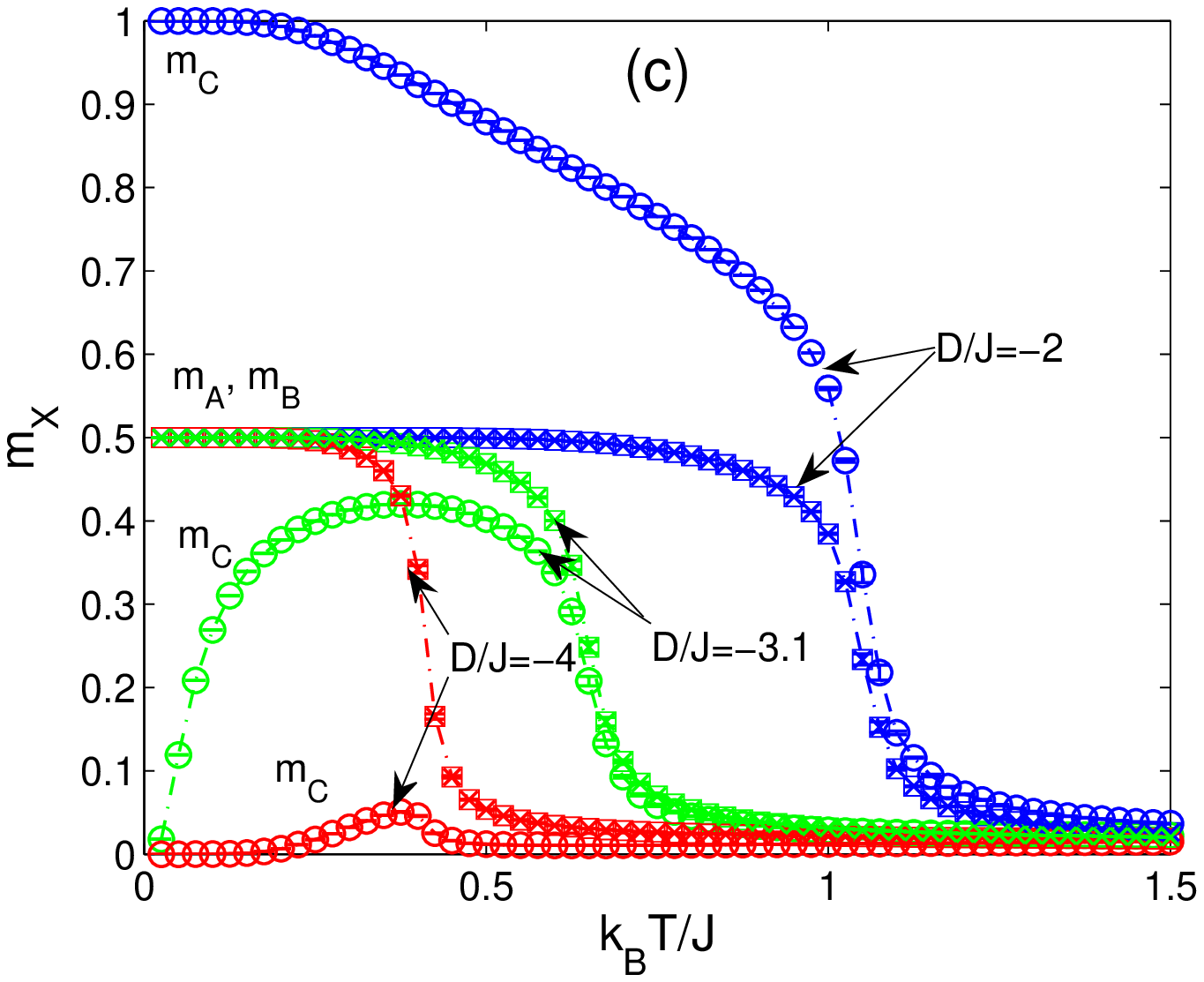}\label{fig:mi-T_L48}}
\caption{Temperature variation of (a) the internal energy and (b) the specific heat and (c) the sublattice magnetizations, for different values of $D/J$ and $L=48$.}\label{fig:e_c_mx-T_L48}
\end{figure}

In Figs.~\ref{fig:e-T_L48} and~\ref{fig:c-T_L48} we plot temperature dependencies of the internal energy and the specific heat, for selected values of the parameter $D/J$, which are below, close to, and above the critical value $D_c/J$. All the internal energy curves in Fig.~\ref{fig:e-T_L48} show some high-temperature anomalies. Those are reflected in the corresponding specific heat dependencies, depicted in Fig.~\ref{fig:c-T_L48}, as pronounced sharp peaks, signifying phase transitions to the low-temperature ferromagnetic states. Furthermore, the energy curves for $D/J=-3.1$ and $-4$ tend to the same value for $T \to 0$, suggesting that the GS phase is $(\pm 1/2,\pm 1/2,0)$, i.e., with the energy independent on the value of $D/J$, as also expected from the above ground-state considerations. 
The types of the respective phases are best understood by looking at the sublattice magnetizations, shown in Fig.~\ref{fig:mi-T_L48}. We can see that for $D/J=-2$ and $-3.1$ all the sublattices start ordering at the same critical temperature but the sublattice magnetizations tend to the ground-state values of $m_{\rm A}=m_{\rm B}=1/2, m_{\rm C}=1$, corresponding to the GS $(\pm 1/2,\pm 1/2,\pm 1)$, only for $D/J=-2$. The curves for $D/J=-3.1$ show finite values below the critical temperature, however, as the temperature is lowered the system enters to the $(\pm 1/2,\pm 1/2,0)$ phase, characterized by $m_{\rm C}$ approaching zero. This phase crossing is also reflected in the second low-temperature peak of the specific heat curve in Fig.~\ref{fig:c-T_L48}. For $D/J=-4$, $m_{\rm C}$ remains zero\footnote{Perfectly zero value can only be achieved for $L \to \infty$} at all temperatures and, thus, the transition from the paramagnetic phase is directly to the phase $(\pm 1/2,\pm 1/2,0)$. Note that in all the cases $m_{\rm A}$ and $m_{\rm B}$ tend to the GS value of 1/2. \\
\begin{figure}[t!]
\centering
\includegraphics[scale=0.48,clip]{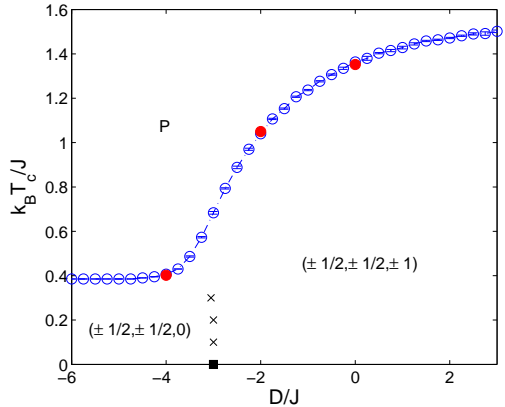}
\caption{Phase diagram of the ${\bf S}=(1/2,1/2,1)$ model in $(k_BT/J-D/J)$ parameter space. The empty circles represent the phase transition temperatures $k_BT_c/J$ between the paramagnetic and long-range order magnetic phases estimated from the specific heat peaks for $L=48$, the filled circles show more precise values obtained from the Binder cumulant crossing, the filled square represents the exact value of the GS transition point $D_c/J=-3$ and the cross-symbols mark positions of the non-critical susceptibility maxima between the magnetic phases.}\label{fig:PD_iftl_mix1}
\end{figure}
\begin{figure}[b!]
\centering
\subfigure{\includegraphics[scale=0.48,clip]{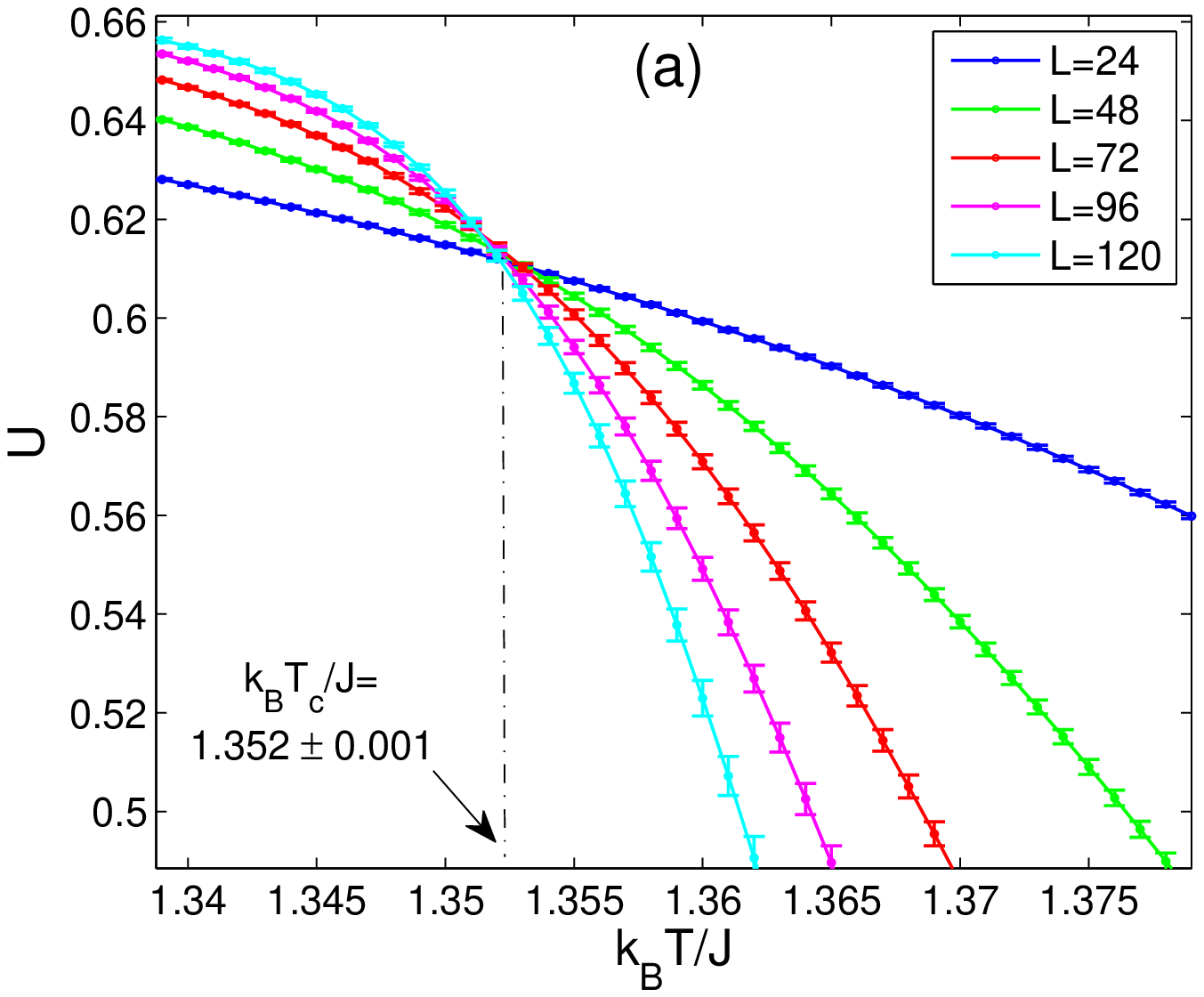}\label{fig:Tc_D0}}
\subfigure{\includegraphics[scale=0.48,clip]{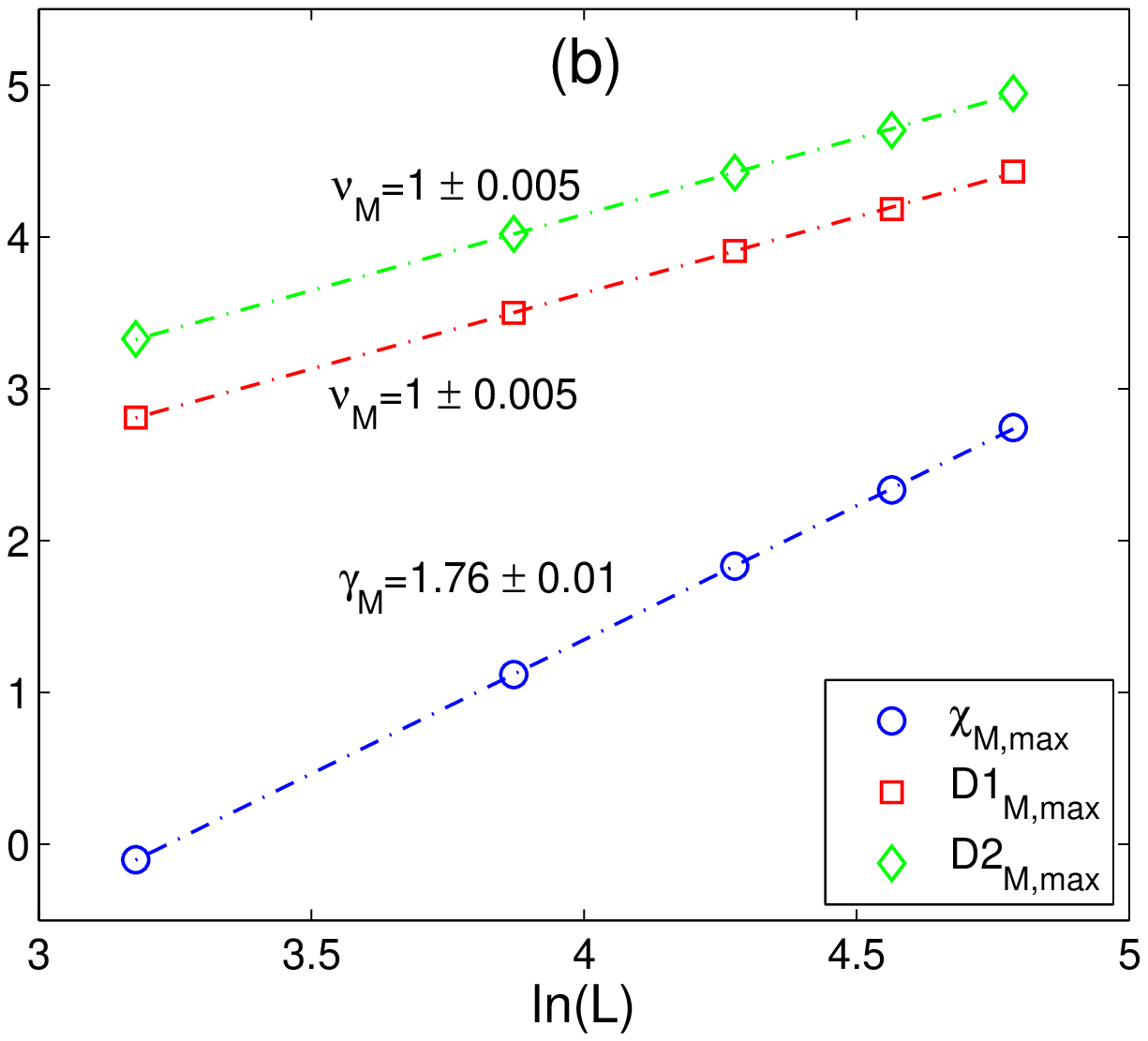}\label{fig:fss_D0}}
\caption{(a) Critical temperature $k_BT_c/J=1.352 \pm 0.001$, for $D/J=0$ obtained from the Binder cumulant crossing method. (b) Correlation length and susceptibility critical exponents $\nu_M$ and $\gamma_M$, respectively, obtained from the FSS analysis for $D/J=0$.}\label{fig:Tc_fss_D0}
\end{figure}
\hspace*{5mm} The resulting phase diagram, estimated from maxima of the specific heat for $L=48$, is presented in Fig.~\ref{fig:PD_iftl_mix1}. More precise values could be obtained from computationally rather intensive FSS analysis or a somewhat simpler method of the Binder cumulant~\cite{bind81} crossing (as an intersection of the Binder parameter
$U$ curves for different lattice sizes $L$). In order to get an idea how well the results obtained for $L=48$ approximate the infinite-limit behavior, we determined the critical temperature at one point for $D/J=0$ by the Binder cumulant crossing method (Fig.~\ref{fig:Tc_D0}). The obtained critical temperature $k_BT_c/J=1.352 \pm 0.001$ at $D/J=0$ is marked in Fig.~\ref{fig:PD_iftl_mix1} by the filled circle and demonstrates that finite effects for $L=48$ only slightly overestimate the critical boundary. Furthermore, the critical value of the Binder cumulant $U(T_c)=0.611$~\cite{kame93} in Fig.~\ref{fig:Tc_D0} as well as the FSS analysis presented in Fig.~\ref{fig:fss_D0} suggest that the transition is second-order and belongs to the standard Ising universality class. We also confirmed that the standard Ising values of the critical exponents are obtained also for some other values than $D/J=0$. Namely, above the critical value $D_c/J=-3$ for $D/J=-2$ we obtained the critical exponents $\nu_M=1.01 \pm 0.01$ and $\gamma_M=1.76 \pm 0.01$ and the critical temperature $k_BT_c/J=1.032 \pm 0.002$ and below $D_c/J=-3$ for $D/J=-4$ we obtained the critical exponents $\nu_M=0.99 \pm 0.01$ and $\gamma_M=1.74 \pm 0.01$ and the critical temperature $k_BT_c/J=0.402 \pm 0.002$. In the limit of $D/J=-\infty$, the model becomes equivalent to the spin-1/2 Ising model on a honeycomb lattice and the critical temperature approaches the exact value of $k_BT_c/J= 0.3797$~\cite{fish67}.\\
\begin{figure}[t!]
\centering
\subfigure{\includegraphics[scale=0.48,clip]{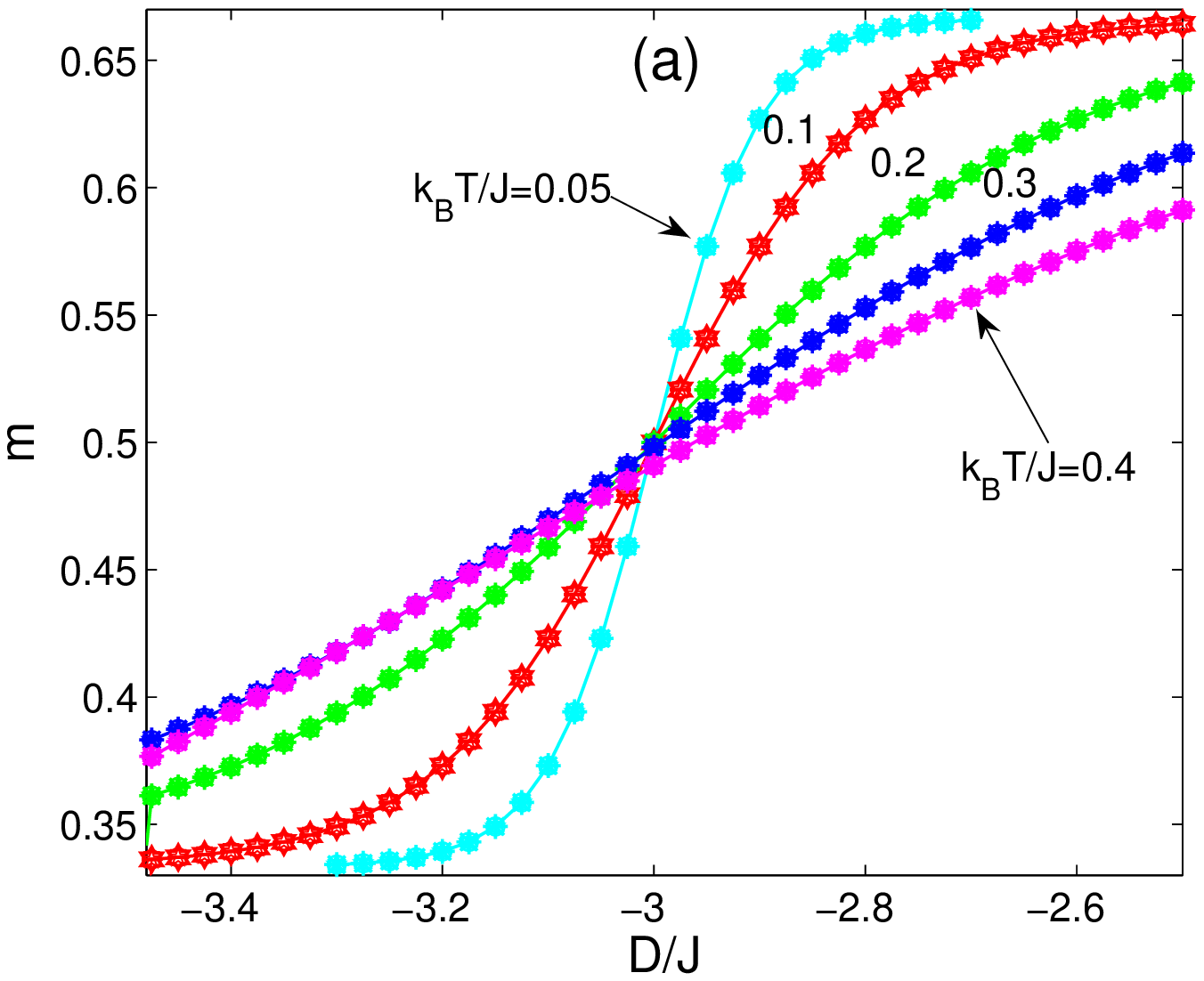}\label{fig:m-D_L24-120}}
\subfigure{\includegraphics[scale=0.48,clip]{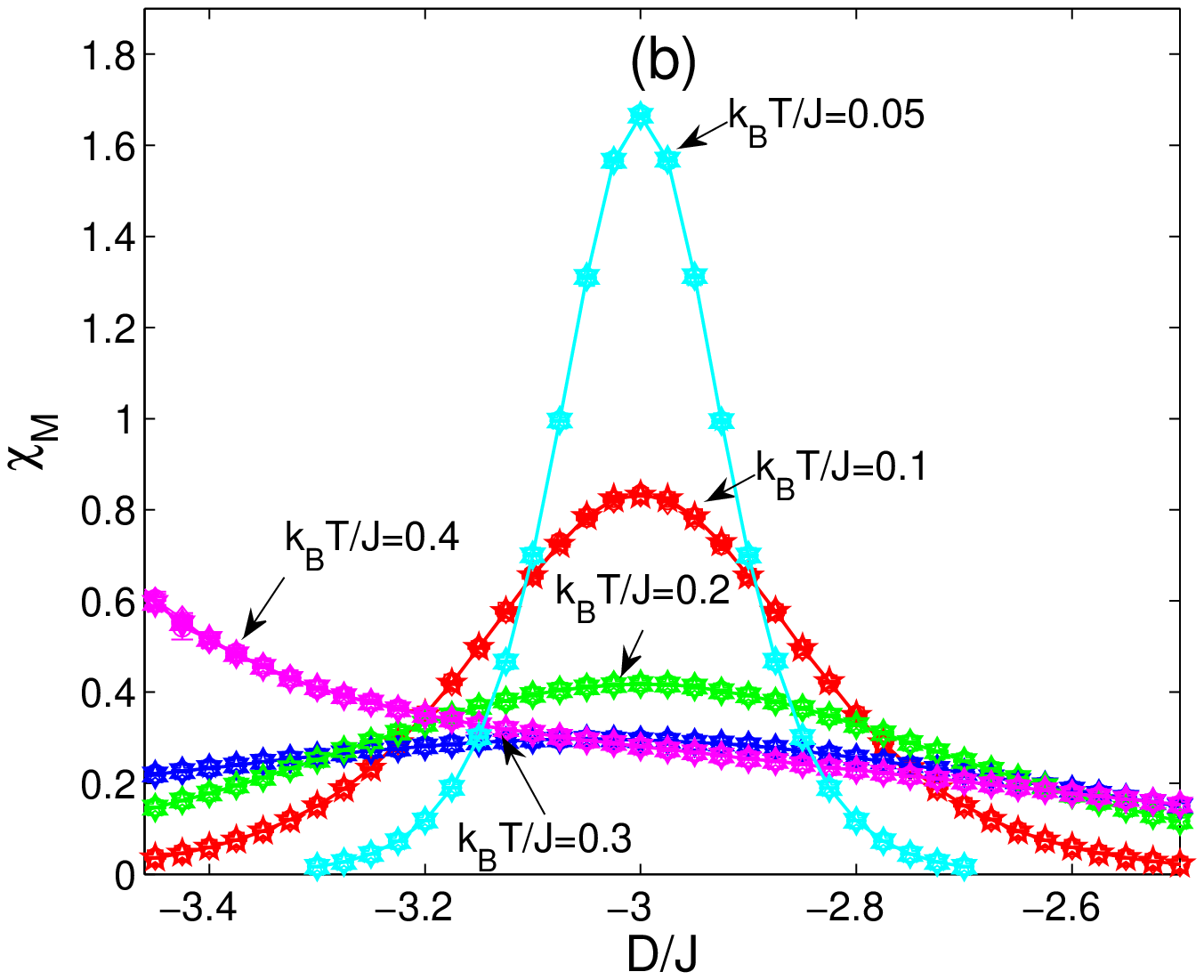}\label{fig:chi-D_L24-120}}
\subfigure{\includegraphics[scale=0.48,clip]{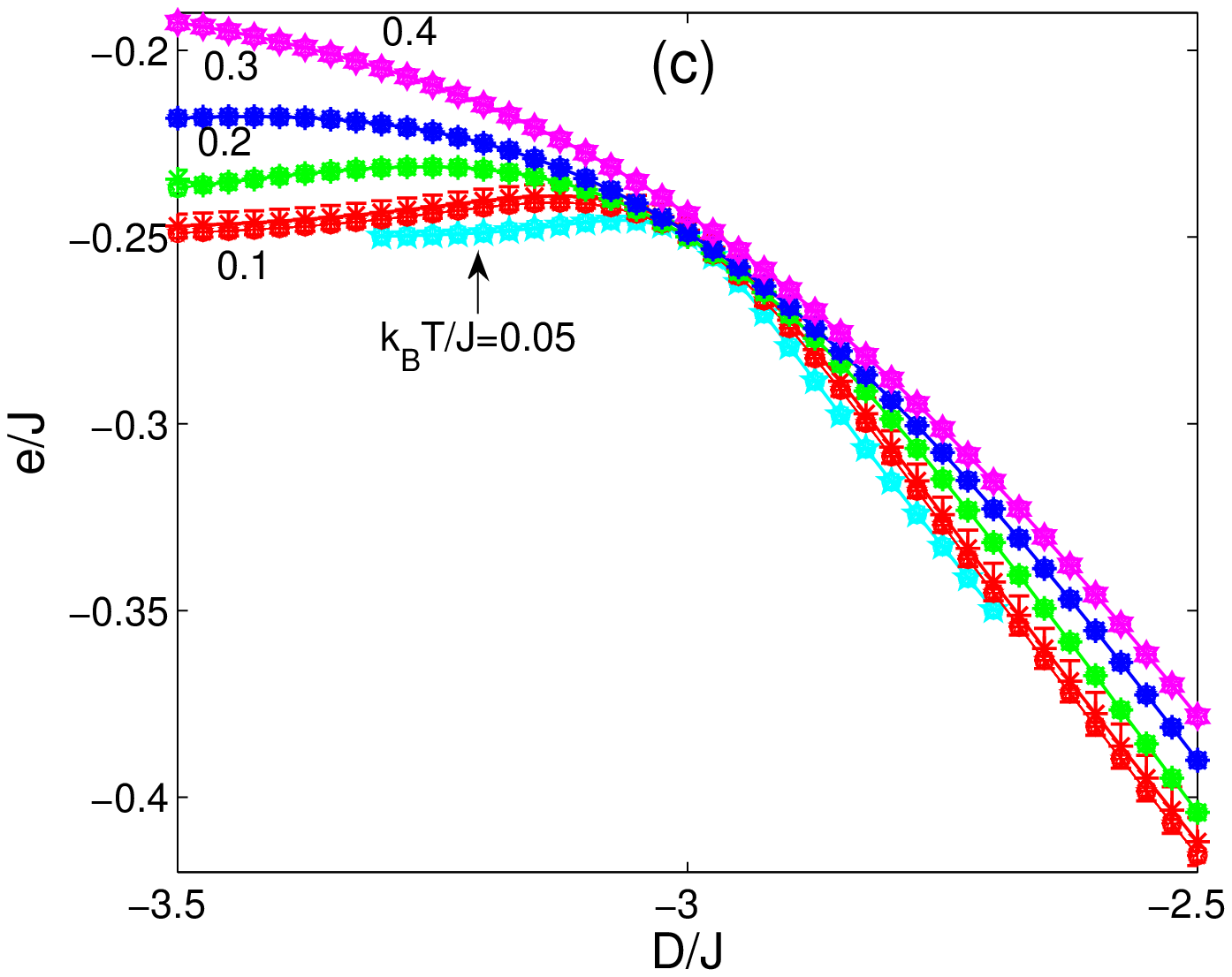}\label{fig:e-D_L24-120}}
\subfigure{\includegraphics[scale=0.48,clip]{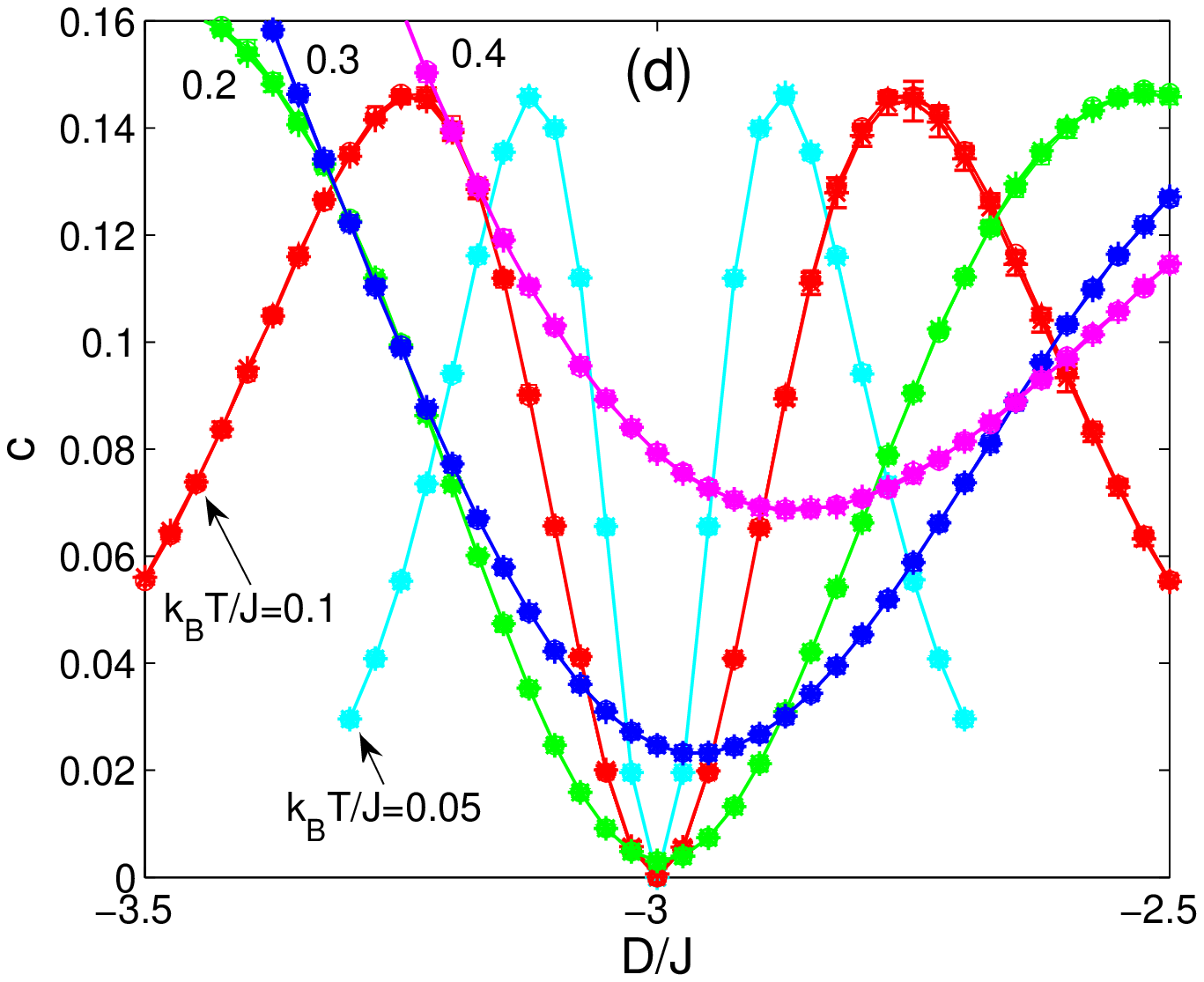}\label{fig:c-D_L24-120}}
\subfigure{\includegraphics[scale=0.48,clip]{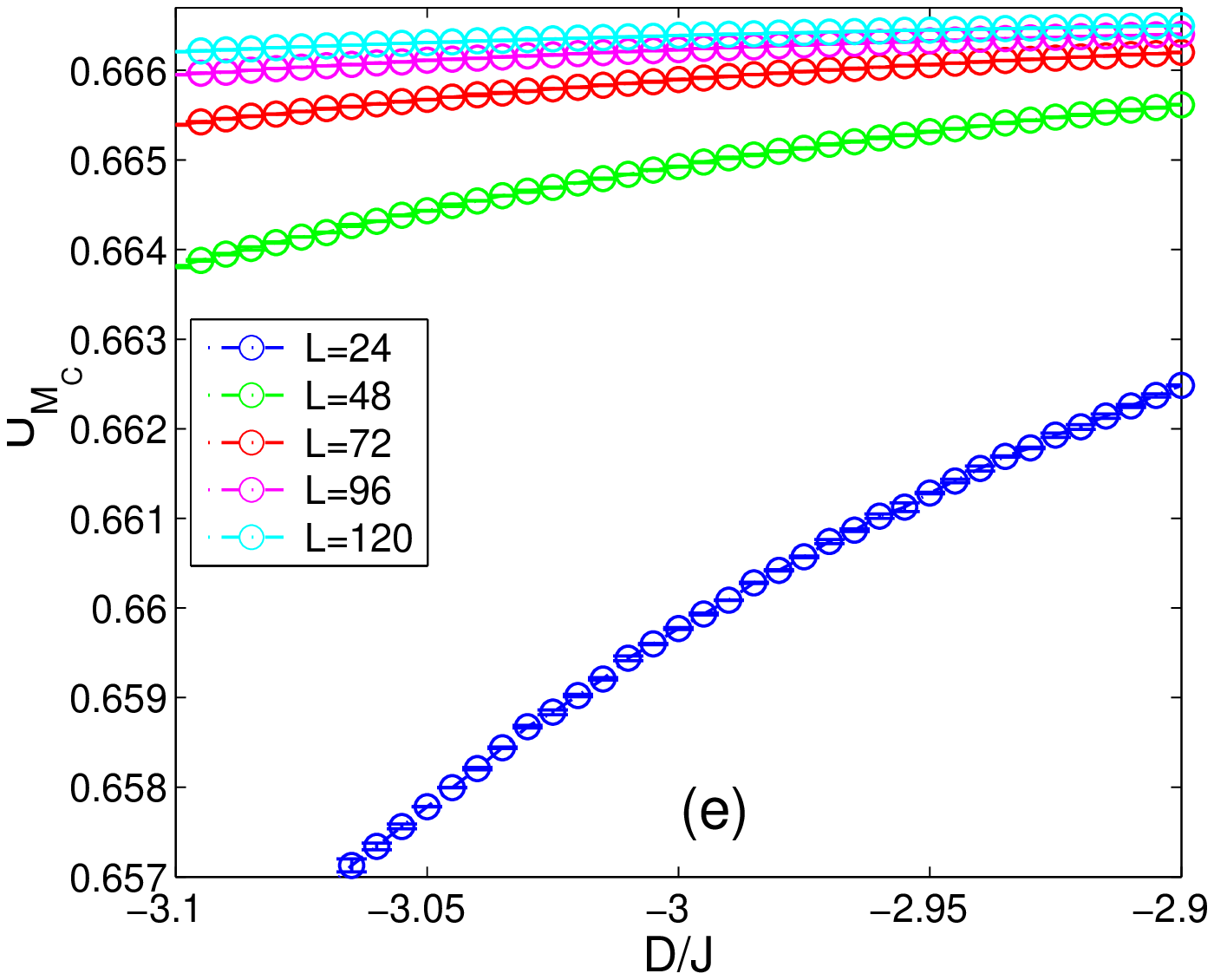}\label{fig:U1-D_L24-120}}
\subfigure{\includegraphics[scale=0.48,clip]{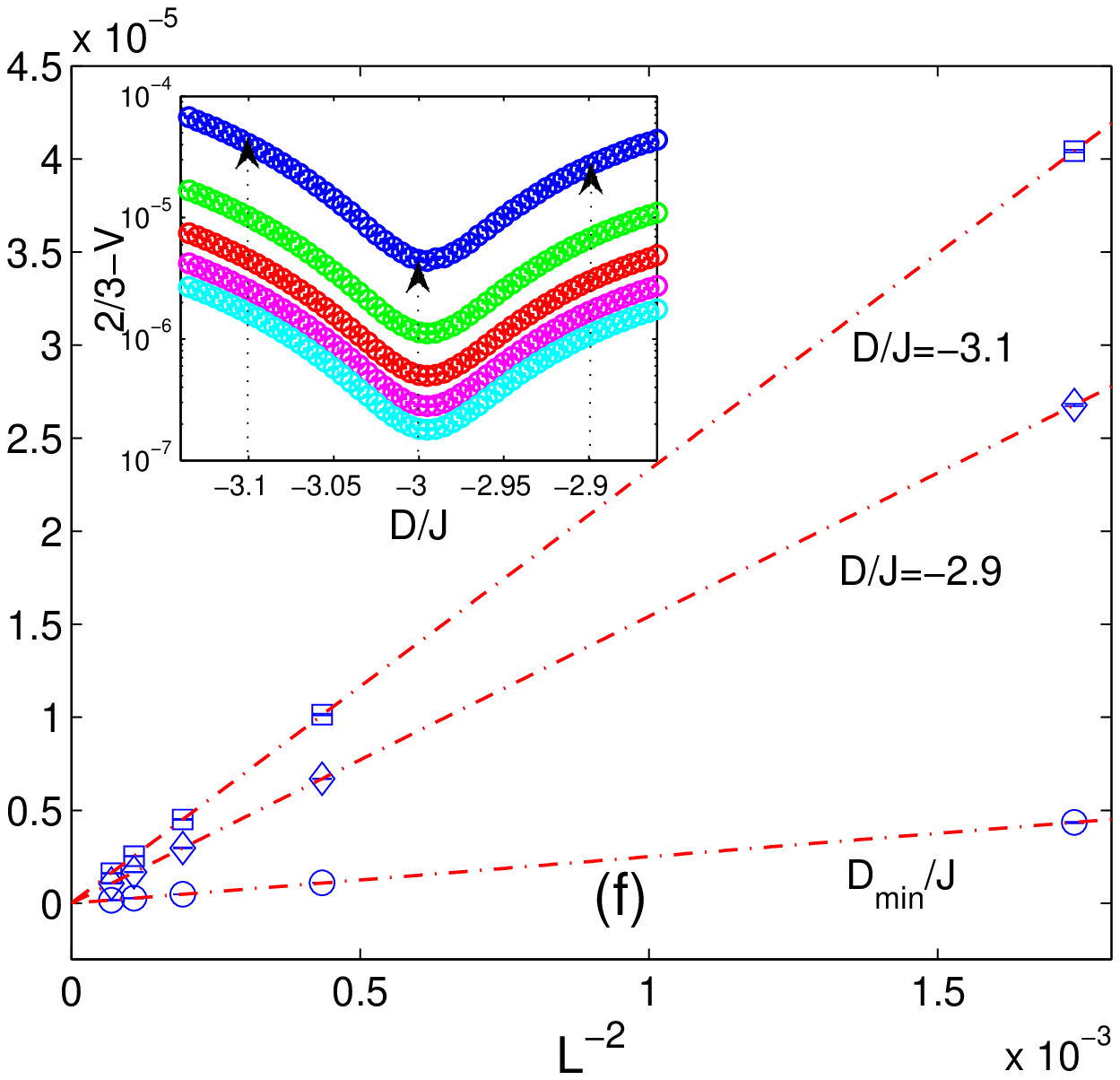}\label{fig:V-D_L24-120}}
\caption{(a) Magnetization, (b) susceptibility, (c) internal energy and (d) specific heat as functions of the parameter $D/J$, for selected low-temperature values. The curves obtained for different lattice sizes $L=24-120$ at each temperature are marked by various symbols but are almost indistinguishable. (e) Forth-order magnetic cumulant $U_{M_{\rm C}}$ and (f-inset) internal energy cumulant $V$ as functions of $D/J$, for $k_BT/J=0.2$ and $L=24-120$. In (f) the FSS analysis of the values of $2/3-V_L(D/J)$ with $L^{-2}$ for $D/J=-3.1$, $-2.9$ and $D_{min}/J$, at which $2/3-V_L(D/J)$ is minimal, gives the following asymptotic values for $L \rightarrow \infty$: $(0.7 \pm 0.5)\times 10^{-8}$, $(0.07 \pm 0.21)\times 10^{-8}$ and $(-0.6 \pm 0.7)\times 10^{-9}$, respectively.}\label{fig:x-D_L24-120}
\end{figure}
\hspace*{5mm} 
As mentioned above, in the ground state the two ordered phases $(\pm 1/2,\pm 1/2,0)$ and $(\pm 1/2,\pm 1/2,\pm 1)$ coexist at the value of $D_c/J=-3$, at which the sublattice magnetization $m_{\rm C}$ shows a jump between 1 and 0. Nevertheless, at finite temperatures no signs of either first- or second-order phase transitions were observed. Namely, by measuring thermodynamic quantities at a fixed temperature as functions of the parameter $D/J$ in the vicinity of the border between the two phases we did observe some anomalies but not such that are characteristic for the first- or second-order phase transitions. In particular, in Fig.~\ref{fig:m-D_L24-120} we can see that the magnetization shows an anomalous change at low temperatures but the corresponding magnetic susceptibility peak heights are not sensitive to the lattice size. This is demonstrated in Fig.~\ref{fig:chi-D_L24-120}, where all the susceptibility curves obtained for various system sizes of $L=24,48,72,96$ and 120 are shown to collapse on the same curve for each temperature. Hence, the cross-symbols plotted in the phase diagram (Fig.~\ref{fig:PD_iftl_mix1}) do not represent phase transition points but just the locations of the non-critical maxima observed in the susceptibility curves. Even more interesting is the behavior of the internal energy and the specific heat. At higher temperatures the former is a decreasing and concave function of $D/J$ and the corresponding specific heat show a minimum close to $D_c/J=-3$, tending to zero for $T \to 0$. However, at sufficiently low temperatures, the internal energy changes to convex just below $D_c/J$ (and even slightly increasing) as well as above $D_c/J$, and the corresponding specific heat curves take the form of a double peak structure, again insensitive to the system size (see Figs.~\ref{fig:e-D_L24-120} and~\ref{fig:c-D_L24-120}). \\
\hspace*{5mm} 
Thus, we believe that the phase transition accompanied with a magnetization jump between the values of $m=1/3$ (phase $(\pm 1/2,\pm 1/2,0)$) and $m=2/3$ (phase $(\pm 1/2,\pm 1/2,\pm 1)$) is limited only to $T=0$. It is difficult to verify by MC simulation if some discontinuity could be observed at extremely low temperatures but we think that at this transition even small thermal fluctuations can smoothen out the GS discontinuity. The reason is that the transition occurs only in sublattice C (A and B are ordered in both phases), which consists of isolated (mutually directly noninteracting) spins. Therefore, the change between magnetic $\pm 1$ and nonmagnetic $0$ states of spins on C-sublattice can be initiated even by subtle thermal fluctuations and as $D/J$ is varied it can proceed gradually without any abrupt collective changes of states of C-sublattice spins. Indeed, the above shown collapse of the curves indicates the presence only short-range correlations and the absence of the diverging correlation length. In order to provide further support for the claim of no first- neither second-order phase transitions we show two more figures (Figs.~\ref{fig:U1-D_L24-120} and \ref{fig:V-D_L24-120}) with the analyses of both the forth-order cumulant corresponding to the order parameter of this transition, i.e. $U_{M_{\rm C}}$, as well as the internal energy $V$, defined by expression (\ref{eq.V}). One can see that the curves of the former for different $L$ do not cross and for $L \to \infty$ tend to the value $2/3$. On the other hand, the energy cumulant in this region $V$ does not show minima but maxima. Moreover, the scaling of the minima of the function $2/3-V$ close to the value $D/J=-3$ does not qualitatively differ from the scaling in other neighboring points, such as at $D/J=-3.1$ or $-2.9$, shown in Fig.~\ref{fig:V-D_L24-120}. Namely, they all scale as $L^{-2}$ but asymptotically tend to $V^*=0$. Thus, no attributes characteristic for either second- or first-order phase transitions are evident. 

\subsubsection{Model ${\bf S}=(1/2,1,1)$}
\begin{figure}[t!]
\centering
\includegraphics[scale=0.48,clip]{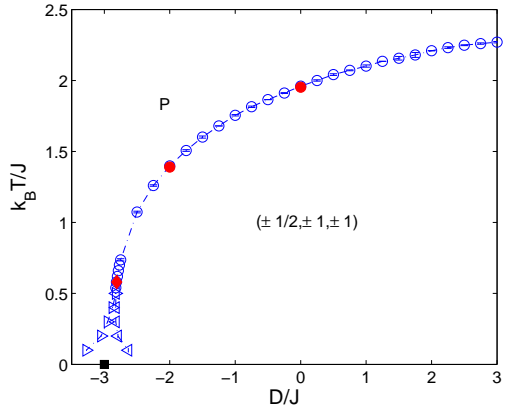}
\caption{Phase diagram of the ${\bf S}=(1/2,1,1)$ model in $(k_BT/J-D/J)$ parameter space. The empty circles represent the phase transition temperatures $k_BT_c/J$ between the paramagnetic and long-range order phase $(\pm 1/2,\pm 1,\pm 1)$ estimated from the specific heat peaks for $L=48$, the filled circles show more precise values obtained from the Binder cumulant crossing, the filled diamond at $(D_t/J,k_BT_t/J) = (-2.81 \pm 0.01,0.58 \pm 0.02)$ is the tricritical point and the filled square represents the exact value of the GS transition point $D_c/J=-3$. The empty triangles mark the hysteresis widths at first-order transitions.}\label{fig:PD_iftl_mix2}
\end{figure}
\begin{figure}[t!]
\centering
\subfigure{\includegraphics[scale=0.48,clip]{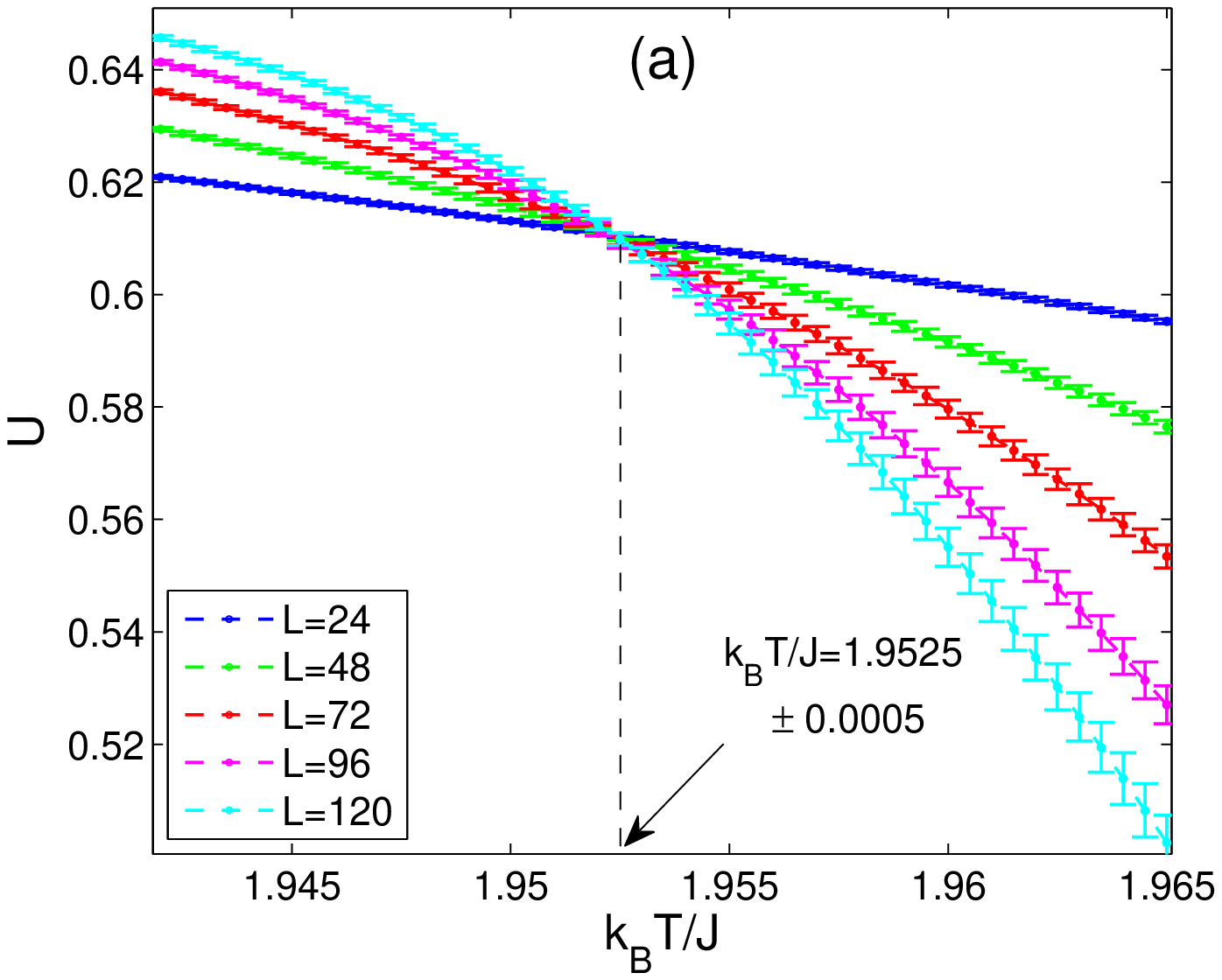}\label{fig:U4_D0a_mix2}}
\subfigure{\includegraphics[scale=0.48,clip]{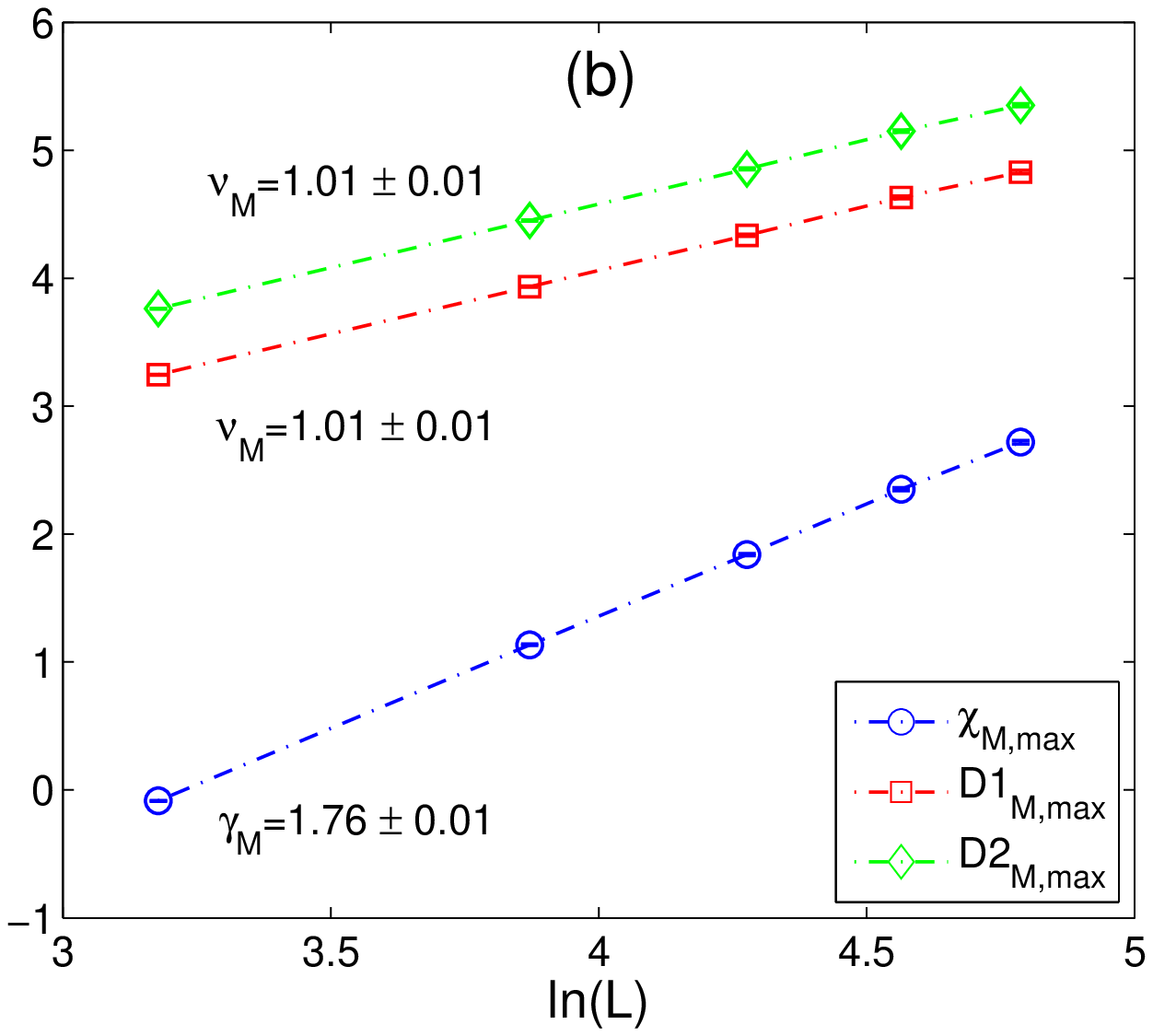}\label{fig:fss_D0_mix2}}
\caption{(a) Critical temperature $k_BT_c/J=1.9525 \pm 0.0005$, for $D/J=0$ obtained from the Binder cumulant crossing method. (b) Correlation length and susceptibility critical exponents $\nu_M$ and $\gamma_M$, respectively, obtained from the finite-size scaling analysis at $D/J=0$.}\label{fig:Tc_fss_D0_mix2}
\end{figure}
\begin{figure}[t!]
\centering
\subfigure{\includegraphics[scale=0.48,clip]{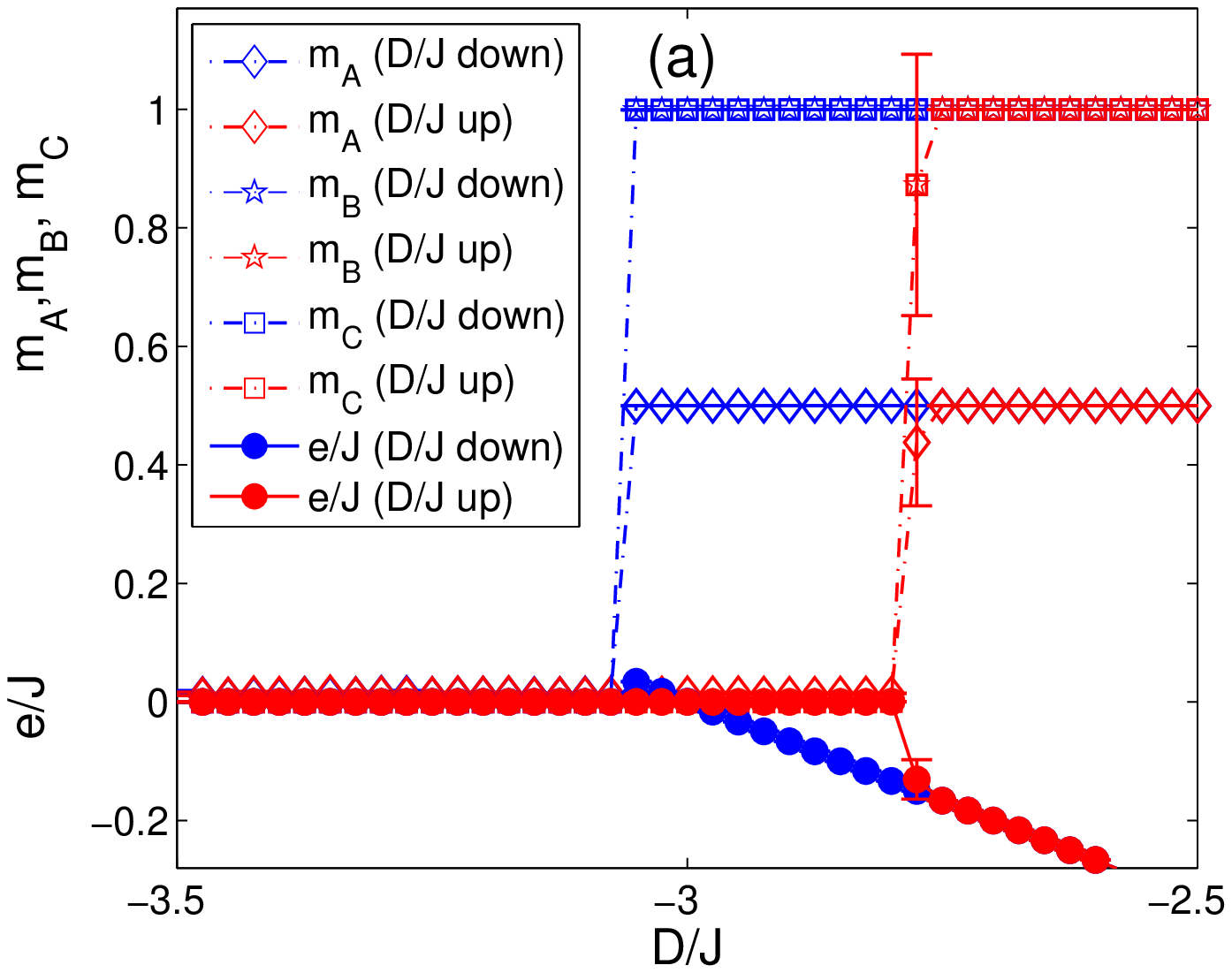}\label{fig:hyst_T02}}
\subfigure{\includegraphics[scale=0.48,clip]{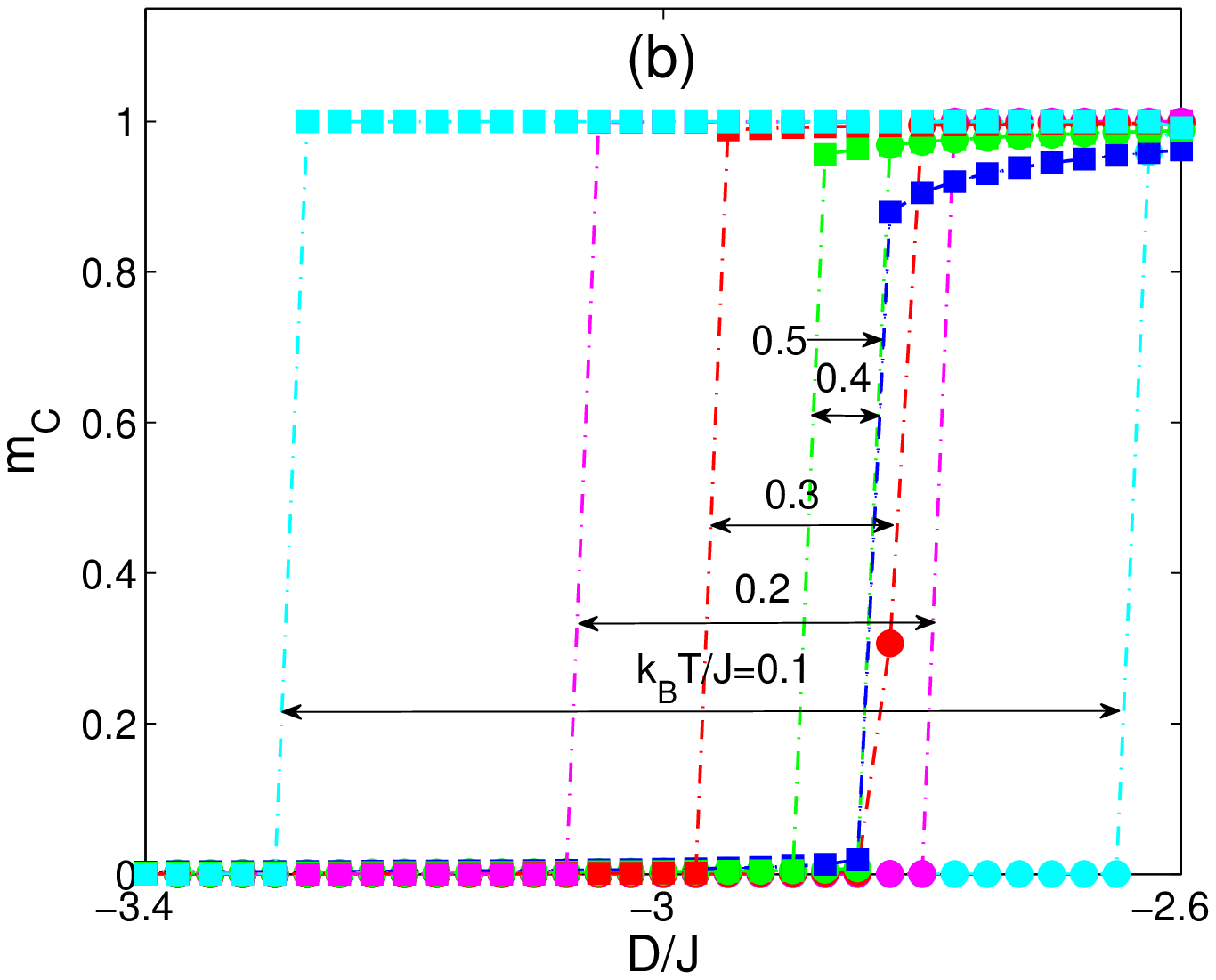}\label{fig:hyst_iftl_mix2}}
\caption{(a) Sublattice magnetizations and internal energy as functions of the parameter $D/J$, for $k_BT/J=0.2$. The blue (red) curves represent the values obtained in the $D/J$ decreasing (increasing) processes. (b) Hysteresis of the sublattice magnetization $m_{\rm C}$ at various temperatures.} \label{fig:hyst}
\end{figure}
\begin{figure}[t!]
\centering
\includegraphics[scale=0.48,clip]{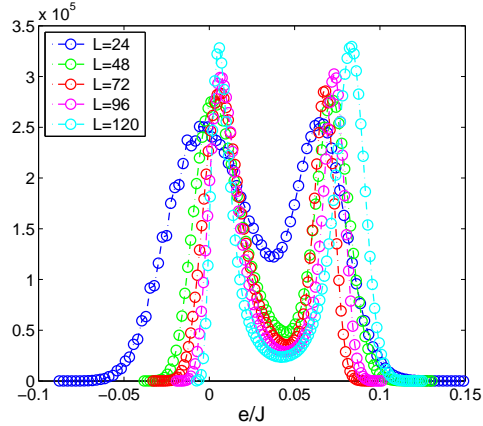}
\caption{Energy histograms collected at $k_BT/J=0.54$ and different values of $D/J$ fine-tuned for each value of $L$ to obtain bimodal distributions with approximately equally high peaks, corresponding to the respective pseudo-transition points.}\label{fig:histogram}
\end{figure}

The phase diagram of the model as a function of the single-ion anisotropy parameter $D/J$, for $D/J$ not too close to the critical value of $D_c/J=-3$, is again estimated from the specific heat maxima for $L=48$ and presented in Fig.~\ref{fig:PD_iftl_mix2}. For $D/J \geq 0$, the phase transition is second-order and belongs to the Ising universality class. This is demonstrated in the FSS analysis performed for $D/J=0$ (Fig.~\ref{fig:fss_D0_mix2}), at which the critical temperature is determined from the Binder cumulant crossing as $k_BT_c/J=1.9525 \pm 0.0005$ and the critical Binder cumulant takes the universal value $U(T_c)=0.611$~\cite{kame93} (Fig.~\ref{fig:U4_D0a_mix2}). \\
\hspace*{5mm} The transition retains the same character also in some range below $D/J=0$. For example, for $D/J=-2$ we also obtained the standard critical exponents $\nu_M=1.00 \pm 0.005$ and $\gamma_M=1.76 \pm 0.01$ and the critical temperature $k_BT_c/J=1.390 \pm 0.005$. However, as the critical GS value of $D/J=-3$ is approached, the critical temperatures rapidly drop and the phase boundary becomes almost vertical. Therefore, in order to locate the critical temperatures in this region, it is more convenient to measure the physical quantities at a fixed temperature as functions of the parameter $D/J$. At low temperatures the behavior of the measured quantities is typical for first-order phase transitions. For example, the sublattice magnetizations and the internal energy show discontinuous and hysteretic behavior, as demonstrated in Fig.~\ref{fig:hyst_T02}, where the anisotropy parameter $D/J$ is decreased and increased at the fixed temperature $k_BT/J=0.2$. At such a low temperature the entropic contribution is almost negligible and the free energy can be fairly well approximated by the internal energy. Then the true transition point can be estimated as a crossing point of the magnetic and paramagnetic energy branches obtained by $D/J$ decreasing and increasing processes, respectively. As expected, the transition point at $k_BT/J=0.2$ is very close to the GS value of $D_c/J=-3$. The blue (red) branches extending below (above) this value represent metastable states. From Fig.~\ref{fig:hyst_iftl_mix2} one can observe that with increasing temperature the hysteresis gets narrower and eventually vanish, which is a sign of the phase transition order change from first to second (tricritical point). From Fig.~\ref{fig:hyst_iftl_mix2} it appears that this happens at $(D_t/J,k_BT_t/J) \approx (-2.84,0.5)$. Nevertheless, by inspecting the magnetization and energy histograms, we found evidence of the discontinuous nature of the transition persisting also above this temperature. The bimodal character of the latter is clearly evident even at $k_BT/J=0.54$. In Fig.~\ref{fig:histogram}, by reweighing of the internal energy histograms to the values of $D/J$ at which peaks of the bimodal distributions achieve approximately equal heights, corresponding to the respective pseudo-transition points, we can observe well separated peaks with a deepening energy barrier between the two phases with increasing system size, a clear characteristic of a first-order transition. A more detailed study on a finer resolution gives a more precise estimate of the tricritical point, at which hysteresis of the quantities completely vanish and their distributions becomes unimodal, as $(D_t/J,k_BT_t/J) = (-2.81 \pm 0.01,0.58 \pm 0.02)$. 

\section{Conclusions}
We have studied the mixed spin-1/2 and spin-1 Ising models on a triangular lattice with sublattices A, B and C, in two mixing modes: $(S_{\rm A},S_{\rm B},S_{\rm C})=(1/2,1/2,1)$ and $(S_{\rm A},S_{\rm B},S_{\rm C})=(1/2,1,1)$. The main goals were to study the nature of the transition between two ferromagnetic phases $(\pm 1/2,\pm 1/2,\pm 1)$ and $(\pm 1/2,\pm 1/2,0)$, expected to appear in the former, and to verify the existence of a tricritical point in the latter. Our Monte Carlo simulations provided evidence of no phase transitions between the phases $(\pm 1/2,\pm 1/2,\pm 1)$ and $(\pm 1/2,\pm 1/2,0)$, at least down to the temperature $k_BT/J=0.05$. This might be attributed to the fact that in the present mixing mode the spin-1 sites are isolated from each other by the spin-1/2 sites residing on the honeycomb backbone (see Fig.~\ref{fig:schem_a}) and, thus, do not cooperate at the transition to the nonmagnetic state. It is interesting to notice that the phase diagram qualitatively resembles that of the spin-3/2 Blume-Capel (BC) model, which in the low-temperature region also gives two distinct ferromagnetic phases $(\pm 1/2,\pm 1/2)$ and $(\pm 3/2,\pm 3/2)$. Nevertheless, in contrast to the present model, in the BC model up to a certain temperature those are separated by a first-order transition line ending in a critical end point~\cite{bekh97,tuck00}. \\
\hspace*{5mm} On the other hand, the mixed-spin ${\bf S}=(1/2,1,1)$ model has been shown to display only one long-range-order state $(\pm 1/2,\pm 1,\pm 1)$. That is separated from the paramagnetic phase by a critical frontier featuring a tricritical point at which a line of second-order phase transition points at higher values of the single-ion anisotropy parameter $D/J$ and higher temperatures meets a line of first-order phase transition points at low temperatures and the anisotropy approaching the value of $D_c/J=-3$ from above. To our best knowledge, this is the only reported example of a two-dimensional mixed spin-1/2 and spin-1 model, which shows a tricritical point. Thus, the present study gives a positive answer to the question regarding the possibility of a tricritical behavior in such two-dimensional mixed-spin systems, providing they are considered on a lattice with sufficiently high coordination number that allows mixing in which spin-1 sites can occupy a connected backbone (such as honeycomb in the present case) network. 
\\
\hspace*{5mm} Finally, we would like to point out that in mixed-spin models on bipartite lattices in zero field the sign of the exchange interaction does not matter and both ferromagnetic and ferrimagnetic interactions will give the same phase diagrams. However, in the present mixed-spin model on a triangular lattice, a negative value of $J$ in the Hamiltonian, either between spins of different (ferrimagnetic) or the same (antiferromagnetic) magnitudes, will bring about geometrical frustration which may substantially change the critical behavior. Such a model is already under investigation and the results will be reported in a separate paper.

\section*{Acknowledgments}
This work was supported by the Scientific Grant Agency of Ministry of Education of Slovak Republic (Grant Nos. 1/0234/12 and 1/0331/15). The authors acknowledge the financial support by the ERDF EU (European Union European Regional Development Fund) grant provided under the contract No. ITMS26220120047 (activity 3.2.).

\end{document}